# The temporal changes in the emission spectrum of Comet 9P/ Tempel 1 after Deep Impact

*William M. Jackson, XueLiang Yang, and Xiaoyu Shi*
*Department of Chemistry, University of California, Davis*
*Anita L. Cochran*
*McDonald Observatory, University of Texas at Austin*

**ABSTRACT**

The time dependence of the changes in the emission spectra of Comet 9P/Tempel 1 after Deep Impact are derived and discussed. This was a unique event because for the first time it gave astronomers the opportunity to follow the time history of the formation and decay of $O(^1S)$, OH, CN, $C_2$, $C_3$, NH, and $NH_2$. Least squares fits of a modified Haser model with constraints using known rate constants were fit to the observed data. In the case of OH a simple two-step Haser model provides a reasonable fit to the observations. Fitting the emissions from $O(^1S)$, CN, $C_2$, $C_3$, NH, and $NH_2$ requires the addition of a delayed component to a regular two or three step Haser model. From this information a picture of the Deep Impact encounter emerges where there is an initial formation of gas and dust, which is responsible for the prompt emission that occurs right after impact. A secondary source of gas starts later after impact when the initial dust has dissipated enough so that solar radiation can reach the surface of freshly exposed material. The implications of this and other results are discussed in terms of the implications on the structure and composition of the comet's nucleus.

## 1. INTRODUCTION

The collision between the Deep Impact projectile and comet 9P/Tempel 1 on 4 July 2005 UT created unique conditions for studying the chemical processes responsible for the radicals that are observed in comets. It is the first time in the history of astronomy that an astronomical event initiated by man could be followed in real time. The energy expended in the collision was $19 \times 10^9$ J. The interaction of the material released in this collision with solar radiation was followed with the Keck I telescope on Mauna Kea using the high-resolution echelle spectrograph (HIRES). Excellent high-resolution spectra of the emissions from $O(^1S)$, OH, CH, CN, $C_2$, $C_3$, NH, and $NH_2$ radicals as a function of time after the collision were measured. All of these emissions were present before the encounter but we have devised a method to separate the emission due to the impact from those that were present before the event and thus as a result have been able to derive the temporal behavior of the emissions. The purpose is to directly determine the lifetime of a particular radical species from the variation of the emission intensity as a function of time. We are not determining scale lengths since we have an independent measurement of time. We employ the time after impact, which is a measured quantity, and only use the distance to separate the emission caused by the impact from the emission present before impact. No velocity is required to change scale length into time since time is measured directly at the telescope. Thus, the data that are extracted from the observations are similar to having a double beam spectrometer in the laboratory that measures the light intensity of the emissions present before impact at the same time as measuring the total light intensity after the impact at each wavelength. By subtracting the former from the later one obtains changes in the emission at each wavelength due to the impact. There have been no previous studies in the history of cometary science like this because there has been no previous time in recorded history where man has initiated an astronomical event and then had the tools to record the response as a function of time after the event.

Models have been developed to fit the temporal behavior of the emissions and thus provide new information about the chemical reactions as well as the properties of the cometary nucleus.

## 2. OBSERVATIONS

The Keck I telescope with the HIRES spectrograph was used to observe the aftermath of the impact. The spectrograph was equipped with the blue cross-disperser which means that it covered a spectral band pass between 3047–5894 Å at a resolution, R = $\lambda/\Delta\lambda$ =47,000. This band pass and resolution allowed us to identify and assign the spectral features of OH, NH, CN, CH, $C_3$, $C_2$, $NH_2$, and O ($^1$S). The extremely good image quality of the optics of the Keck telescope and the stable atmosphere resulted in seeing of 0.7 arcsec. Thus, by employing a slit size of 7.0 x 0.86 arcsec (or 4570 x 562 km at the comet) we were able to obtain excellent spatial resolution of ~ 457 km at the comet. This kind of spatial resolution allowed us to separate the temporal response of these emissions from the ambient emission already present in the coma of the comet.

The observations on 4 July 2005 started during nautical twilight, at 05:36:15, so that we could obtain a pre-impact spectrum. Eight degree twilight was at 05:29UT; 12 degree twilight (nautical) was at 05:58UT; 18 degree twilight (astronomical) was at 06:28UT. Complete details of the observations as well as of our preliminary reduction, where we extracted the integrated spectra over the entire slit length as well as only in the inner 0.7 arcsec, are detailed in Cochran et al. (2007). A log of the observations is given in that paper.

The impact caused the release of additional gas from the nucleus beyond the normal outflow of ambient gas. In order to understand this additional gas, it is necessary to remove the ambient gas signal from the post-impact observations. It was only after using the excellent spatial and spectral resolution of the Keck-HIRES that we were able to obtain the unique chemical signatures of the Deep Impact event. The impact caused an instantaneous release of gas followed by that gas flowing outwards from the impact site. In order to study the lifetimes of the molecules against photodissociation, we wanted to derive light curves that included *only* the gas produced by the impact with none of the ambient gas signature. Because the gas flowed outward at some finite speed (we assumed 0.55 km sec$^{-1}$ but as we will show later in this paper, this is an upper limit to gas velocity.), it did not reach the end of our 7 arcsec slit until after our fifth observation post impact. Thus, the ends of our slit during these first observations would still be an accurate and concurrent measure of the ambient cometary spectrum. We therefore extracted the spectra from the ends of the slit, averaging over the spectra from both ends of the slit from the first four spectra, and defined this spectrum to be the ambient cometary spectrum. This spectrum did not contain any of the gas or dust triggered in the impact. Once the ambient spectrum was obtained in this manner, we extracted the impact spectrum by using an adaptive aperture, which was sized to follow the outflow of gas from the aperture. Thus, in the first 10-minute spectrum, the gas had flowed 273 km from the nucleus, or 1.7 pixels, and we would extract a spectrum ±1.7 pixels from the optocenter. The ambient spectrum was then removed from the spectrum extracted over this adaptive aperture size, leaving a spectrum that only contained the gas resulting from the impact. By the 6$^{th}$ spectrum, the impact material had filled the aperture and we extracted the spectrum over the whole slit from that point onwards. After the material filled the slit, the new impact material would be flowing out of the slit, as will be discussed below when we derive the model. Jackson and Cochran (2000) provided a detailed description of this reduction procedure in an earlier paper (see Table 1 of that paper for the number of pixels traveled for each spectrum).

The cometary ambient gas signal is also modulated by the rotation and the amount of change is different for each species. This modulation is small compared to the signal because the rotation period of the nucleus is of order 41 hours (1.7 days) with a broadband light modulation at maximum of 0.5 magnitudes (Lamy et al. 2007). The modulation amplitude of the coma due to the rotation is substantially smaller than 0.5 magnitudes since the observations described in this paper were obtained over only a very small fraction of the 41-hour rotation period. Indeed, the ambient spectrum did not change within our measurement accuracy during the 4 (45 minutes) observations over which we obtained the ambient spectrum from the ends of the slit. The impact signal was very much larger than any change to be expected in the ambient spectrum due to rotation during the time of our observations and it is highly unlikely that there is an outburst with the equivalent energy in the same amount of time as Deep Impact. Thus, we ignored any possible changes due to rotation.

The procedure, which we discuss above and in greater detail in Jackson and Cochran (2008), differs substantially from the approach taken in Manfroid et al. (2007) and Cochran et al. (2007). In those papers, the data were extracted along the full slit; thus they sampled the increased gas from the impact *along with* the ambient spectrum, i.e., the impact signal was a delta on the ambient signal. However, in studying the changes just due to the impact, we wanted to remove the normal cometary activity. Since the spectrum just prior to the start of the impact was obtained with the sky relatively bright (they were started at civil twilight), the contribution of the Earth's skylight makes the pre-impact spectrum useless to remove the ambient activity. In the case of Cochran et al. (2007), we attempted to minimize this effect by also extracting spectra over only the inner 3 pixels to see the impact. This concept does not allow for the outflow of material from the slit so does not produce the true signature of the impact. Manfroid et al.(2007) did not try to remove the normal cometary activity since they were studying the longer-term trend and just looked at the signal above that trend by extending the HIRES data with UVES data. For their purposes, this was sufficient. However, we desired to look solely at the change in chemistry of the coma from the instantaneous impact. This is why we used the approach outlined in Jackson and Cochran (2008). This also explains the differences in the "light curves" between those works and this one since we are showing just the delta due to the impact. Indeed, the curves shown in Manfroid et al. and Cochran et al. are very similar and any differences are the result of the handling of the extra-scattered light discussed in Cochran et al. and choices of continuum removal and the band passes for the integration. In the case of those papers, the impulse signal is distorted by the inclusion of the ambient spectrum in the first five spectra, which dilutes the true signal of just the impact.

## 3. MATHEMATICAL MODEL

To model the observations, we started with the derivation of the Haser model (Haser, 1957), which was given in Manfroid et al. (2007). However, we found that their parameterization was insufficient to describe the data. Instead, we needed to add a delayed contribution to many of the species. Thus, we derived a new parameterization of the model that is described here. This new model for some of the species includes a driving function that has to be incorporated into the model to account for sources of radicals emitted from the ice after the collision. This driving function has a characteristic time for the emission of that species and that tends to be nearly the same for all emitters. The same slit width, slit length, and flow velocity were used to fit all of the curves for the different cometary emissions and only the rate constants and the mechanisms were changed to fit each of the curves. The model is symmetric with respect to the parent and daughter species with no preference for $k_G$ or $k_D$ to be the smaller value. In that sense, the model is degenerate (see for example Cochran and Schleicher 1993). However, additional constraints from laboratory and theoretical work, as well as, astrophysical considerations have led us to choose the values listed here. We have sought to keep the model as simple as possible and have chosen the least number of variables required to fit the observed data. Following an outline of the equations used in the model, a brief discussion of the temporal response of each emission will be presented to illustrate the information that is obtained when we model the data.

The most elaborate form of the model employs three steps to produce the observed free radical. In this case a grandparent produces a parent, which then produces the daughter that is observed in the observations. The grandparent was initially generated in the gas phase, along with the dust, by the impact of the spacecraft with the comet at time t = 0. The assumption is that there is no continuous generation of the grandparent beyond what is already present in the absence of the spacecraft colliding with the comet. This background that is extracted pixel by pixel has already been removed from the data. The *g* factors of the emissions are not included in the model because a steady state of the excited state involved for each of them is quickly reached, so it does not affect the concentration of the ground state of the molecule. In this case the reaction can be simply written as three-step model:

$$Grandparent \rightarrow Parent \quad (1)$$
$$Parent \rightarrow Daughter \quad (2)$$
$$Daughter \rightarrow Product \quad (3)$$

Within this framework, we may write:

$$\frac{dn_G}{dt} = -k_G n_G \qquad (4)$$

$$\frac{dn_P}{dt} = \alpha_{GP} k_G n_G - k_P n_P \qquad (5)$$

$$\frac{dn_D}{dt} = k_P \alpha_{PD} n_P - k_D n_D \qquad (6)$$

The $n_j$ is the number density of $j$ molecules and $k_j$ is the rate constant in (s$^{-1}$) for the $j^{th}$ molecule. In equations 4 through 6 the $\alpha_{GP}$ and the $\alpha_{PD}$ refer to the branching ratio for grandparent or parent producing a specific parent or daughter, whereas $k_G$, $k_P$, and $k_D$ correspond to the rate constants for the total loss of the grandparent, parent or daughter to all channels.

Solution of a daughter derived from Eqns. 4, 5, and 6
with the boundary conditions (BC) at t = 0 of $n_G = n_G^0$, $n_P = 0$, and $n_D = 0$ give

$$n_D = n_G^0 \alpha_{PD} \alpha_{GP} \left[ \frac{k_G k_P \left(e^{-k_G t} - e^{-k_D t}\right)}{(k_G - k_P)(k_G - k_D)} + \frac{k_G k_P \left(e^{-k_D t} - e^{-k_P t}\right)}{(k_G - k_P)(k_P - k_D)} \right] \qquad (7)$$

$$n_D = n_G^0 \alpha_{PD} \alpha_{GP} R(t)$$

*where* R(t) is the time dependence of the daughter production.

After the impact, solar radiation will strike the newly exposed surface of the comet and this presents a new source of gas for the coma. As a result Eqn. (4) will change to Eqn. (8) and this new set of differential equations will require a new solution shown in Eqn. (9).

$$\frac{dn_G}{dt} + k_G n_G = \gamma k_i e^{-k_i t} \qquad (8)$$

A new quantity, γ, is introduced in Eqn. (8) that represents the original surface density of the grandparent from the newly exposed icy surface of the comet. This γ, represents the sublimation rate, $E$, the number of molecules cm$^{-2}$ s$^{-1}$ divided by the velocity of the gas coming of the surface. The $k_i$, is interpreted as the rate constant for the surface density to be depleted to 1/e of its original value. The solution of these new equations is now given in equation (9).

The solution with the boundary conditions at t = 0 of $n_P = 0$ and $n_D = 0$ is

$$n_{Di} = \gamma \alpha_{PD} \alpha_{GP} \left[ \frac{k_i k_G k_P \left(e^{-k_D t} - e^{-k_P t}\right)}{(k_i - k_G)(k_G - k_P)(k_P - k_D)} + \frac{k_i k_G k_P \left(e^{-k_G t} - e^{-k_D t}\right)}{(k_i - k_G)(k_G - k_P)(k_G - k_D)} \right.$$
$$\left. + \frac{k_i k_G k_P \left(e^{-k_P t} - e^{-k_D t}\right)}{(k_i - k_G)(k_i - k_P)(k_P - k_D)} + \frac{k_i k_G k_P \left(e^{-k_D t} - e^{-k_i t}\right)}{(k_i - k_G)(k_i - k_P)(k_i - k_D)} \right] \quad (9)$$
$$= \gamma \alpha_{PD} \alpha_{GP} R(t)_i$$

This equation is used when the observed data points indicate there is a need for including in the time dependence, $R(t)_i$, an additional source of gas sublimating from the ice. The time dependence of the data is then just the sum of $n_D + n_{Di}$. Since we are interested only in comparing the temporal data to the photodissociation rate constants for the grandparents, parents, and daughters we normalized all of the data to the maximum value in each plot. This reduces the above equations to the following:

$$n_D + n_{Di} = n_G^0 \alpha_{PD} \alpha_{GP} \left[ \frac{k_G k_P \left(e^{-k_G t} - e^{-k_D t}\right)}{(k_G - k_P)(k_G - k_D)} + \frac{k_G k_P \left(e^{-k_D t} - e^{-k_P t}\right)}{(k_G - k_P)(k_P - k_D)} \right]$$

$$+ \gamma \alpha_{PD} \alpha_{GP} \left[ \frac{k_P k_i k_G \left(e^{-k_D t} - e^{-k_P t}\right)}{(k_i - k_G)(k_G - k_P)(k_P - k_D)} + \frac{k_P k_i k_G \left(e^{-k_G t} - e^{-k_D t}\right)}{(k_i - k_G)(k_G - k_P)(k_G - k_D)} \right.$$
$$\left. + \frac{k_P k_i k_G \left(e^{-k_P t} - e^{-k_D t}\right)}{(k_i - k_G)(k_i - k_P)(k_P - k_D)} + \frac{k_P k_i k_G \left(e^{-k_D t} - e^{-k_i t}\right)}{(k_i - k_G)(k_i - k_P)(k_i - k_D)} \right]$$

$$n_D + n_{Di} = n_P^0 \alpha_{PD} \alpha_{GP} R(t) + \gamma \alpha_{PD} \alpha_{GP} R(t)_i$$

$$\frac{n_D + n_{Di}}{\gamma \alpha_{PD} \alpha_{GP}} = \frac{n_P^0}{\gamma} R(t) + R(t)_i = \Phi(t) \quad (10)$$

If there is no grandparent then the prompt process can be written as a two-step model:

$$\frac{dn_P}{dt} = -k_P n_P dt; \frac{dn_D}{dt} = \alpha_{PD} k_P n_P - k_D n_D$$

Th solution with the boundary conditions at t = 0 of $n_P = n_P^0$ and $n_D = 0$ is

$$n_D = n_P^0 \alpha_{PD} \left[ \frac{k_P (e^{-k_D t} - e^{-k_P t})}{(k_P - k_D)} \right]$$

$$n_D = n_P^0 \alpha_{PD} R(t) \quad (11)$$

The delayed process for a two-step model can be derived in a manner similar to the three-step model above to yield:

$$n_{Di} = \gamma\alpha_{PD}\left[\frac{k_i k_P \left(e^{-k_P t} - e^{-k_D t}\right)}{(k_i - k_P)(k_D - k_P)} - \frac{k_i k_P \left(e^{-k_i t} - e^{-k_D t}\right)}{(k_i - k_P)(k_D - k_i)}\right] \quad (12)$$

$$n_{Di} = \gamma\alpha_{PD} R(t)$$

Again, when both prompt and delayed steps have to be used to fit the observed data we have to sum them to obtain the following relationship:

$$n_D = n_P^0 \alpha_{PD} R(t) + \gamma\alpha_{PD} R(t)_i$$

$$\frac{n_D}{\gamma\alpha_{PD}} = \frac{n_P^0}{\gamma} R(t) + R(t)_i = \Phi(t) \quad (13)$$

To incorporate the fact that these equations are only valid when the gas is within the slit we define a function $S(v,t)$ using the slit width, $t_w$ and the slit length, $t_l$ as before:

$$S(v,t) = 1 \qquad t \leq t_w$$

$$S(v,t) = \frac{t_w}{t} \qquad t_w \leq t \leq t_l \quad (14)$$

$$S(v,t) = \frac{2}{\pi}\frac{t_w}{t}\arcsin\frac{t_l}{t} \qquad t_l \leq t$$

Since the gas does not have a single velocity we introduce a velocity distribution, $D(v)$, and then integrate as before to obtain an expression for the time dependence of the daughter distribution at the telescope, $n'(t)$:

$$n'_d(t) = \int N_{go} D(v) \Phi(t) S(v,t) dv \quad (15)$$

This integral is broken up into three different temporal regions, corresponding to $0 - t_w$, $t_w - t_l$, and $t \geq t_l$. The result of the integration on the right hand side of this equation of $D(v)$, $\Phi(t)$, and $S(v,t)$ is used to evaluate $n'_d(t)$. The resulting equation is used to fit the time response of the individual emissions observed during Deep Impact. This fitting is done with a least squares program. In the program at a given time a value $[n^0_p]/[\gamma]$ is chosen and then all of the rate constants are varied to minimize the error, $\Delta^2$, between the model and the experimental data points. The quantity $\Delta^2$ is defined as $\Sigma_j \{I_j(\text{model}) - I_j(\text{experimental})\}^2$, where $I_j \equiv$ intensity at time J. In the two-step model the $k_P$ and $k_D$ are varied in two loops with one nested inside the other. A value of $k_P$ is chosen and then the $k_D$ is varied over the whole range of values to obtain the value that minimizes the differences between the intensity, $I_j(\text{model})$ and the intensity, $I_j$ (experimental) at time j. This process is repeated every 100 s to cover the observation period to obtain a minimum value of $\Delta^2$ and to generate the theoretical curves shown in the plots. For the three step models a similar procedure is used involving three nested loops.

The time dependence of a daughter species is obtained by first correcting the spectra for the emission from the species that are already present before impact as described earlier in this paper and in more detail in our previous paper. The area under the lines in the corrected spectra are then integrated and summed to obtain the total contribution from a radical species at a given time after impact. These then give the data points that are used to compare with the model that is given above.

In the next sections the results of the comparison of the model with the observations will be discussed. The goal is to determine the mechanism for the production of each of the observed radicals and to determine the rate constants associated with the mechanism. In this way we hope to use the results to identify the parents of the observed radical species and additional characteristics of the comet.

## 4. INDIVIDUAL EMISSIONS

**OH**

The relative time response of the OH radical is given in Fig. 1 and it is fitted with a curve based upon a two-step model that involves photodissociation of $H_2O$ to produce OH and then photodissociation of the OH radical. To obtain this curve the rate constant for the photodissociation of OH, $k_D$, was fixed using the literature value and then the rate constant for the photodissociation of $H_2O$ was varied. The literature value of $k_d(OH)$ is $5.6 \times 10^{-6}$ $s^{-1}$ at 1 AU (Singh et al.1983; Schleicher and A'Hearn 1988; van Dishoeck and

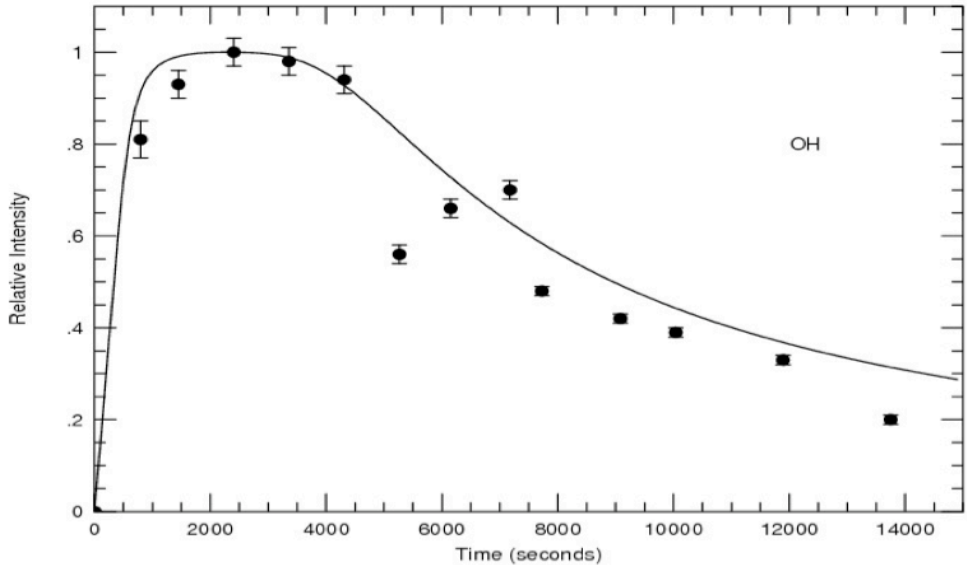

Figure 1. Least squares fit to the OH data using a prompt and delayed two-step model. The error bars are 5 σ and they were computed from the statistical error associated with total number of photons measured at each time. The weight of the delayed model was zero.

Dalgarno 1984), calculated using the measured $A^2\Sigma^- \rightarrow X^2\Pi$ radiative lifetimes along with the modifications by van Dishoeck and Dalgarno 1984. The radiative lifetime can be precisely determined in the laboratory because it only requires the use of the relative decrease in intensity of the emission of radiative lifetime of the individual rotational levels in the v" =0,1, and 2 levels. This then allows one to determine the rate constants for predissociation from the changes observed in these lifetimes as a function of wavelength. The impact occurred when the comet was at 1.51 AU, which decreases the value to $2.5 \times 10^{-6}$ s$^{-1}$. The best fit that could be obtained to the data with the constraints that both rate constants have to be in reasonable agreement with the laboratory data for photodissociation of OH and $H_2O$ yields $k_D$ of $4 \times 10^{-6}$ s$^{-1}$ for OH and a $k_P$ of $2 \times 10^{-5}$ s$^{-1}$ with a $\Delta^2$ equal to 0.13. Lowering the value for $k_D$ to $2 \times 10^{-6}$ s$^{-1}$ results in a $\Delta^2 = 0.16$ and a $k_P$ $0.5 \times 10^{-5}$ s$^{-1}$ value but it puts the value of $k_P$ at the lower limit of literature values for this quantity reported by Crovisier (1994) of 0.5 to $2 \times 10^{-5}$ s$^{-1}$. Since the overall fit is not as good as the one obtain with the previous value we have chosen the former values with the lowest $\Delta^2$. Adding a two-step delayed component and searching the parameter space for a value for the relative contribution of the delayed component leads to a value of zero for this contribution. The present observations and modeling are in reasonable agreement with other evidence that water is the principal molecule produced in comets and that photodestruction is responsible for the production of most of the OH radicals and H atoms that are observed in the emission spectra of comets (Huebner 1990; Brandt and Chapman 2004) via reactions 16 and 17.

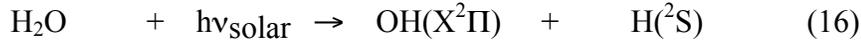

$H_2O \quad + \quad h\nu_{solar} \quad \rightarrow \quad OH(X^2\Pi) \quad + \quad H(^2S) \quad\quad (16)$

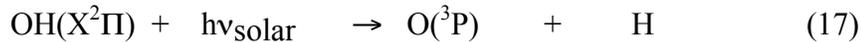

$OH(X^2\Pi) + \quad h\nu_{solar} \quad \rightarrow \quad O(^3P) \quad + \quad H \quad\quad (17)$

The fit to the data shown in Fig. 1 gives a global view of the time dependence and provides evidence that the rate constants derived from the modeling are better than a factor of 2 to 4 of the true values. The modeling in this particular case does have several problems. The fit is not as good as the 5$\sigma$ error bars, it does not fit the fine structure in the data and the rate constants are larger than expected. It is clear that the differential equations derived from the modified Haser model cannot fit the fine structure in the points. We also know that this model ignores such things as opacity of the coma, differential velocities, interactions of dust with gas, electron interaction with the gas, etc. These results suggest that these details need to be added to fit any fine structure observed in the data.

## $O(^1S)$

There are several constraints that must be used in fitting the green emission from $O(^1S)$. This and the red emission are the only prompt electronic emissions observed in comets. Thus, any model for these emissions has to include the known rate constant for the emission from $O(^1S) \rightarrow O(^1D)$ that is fast compared to the other rate constants used for modeling. The red and green emissions are observed in other comets so one suspects that the most dominant oxygen-containing molecule in comets, $H_2O$, must be involved in their production. To date, laboratory experiments have confirmed that $O(^1D)$ is produced during the photolysis of water but not $O(^1S)$ (Huestis 2006).

Examination of the time response of the green emission in Fig. 2 shows that the data has a shelf at the longer times. This kind of behavior cannot be modeled with a simple two or a three-step model but requires the introduction of the exponential driving function with another two or three-step model. In fact, as will be shown, all of the emissions that resulted from the Deep Impact encounter other than OH require such a delay.

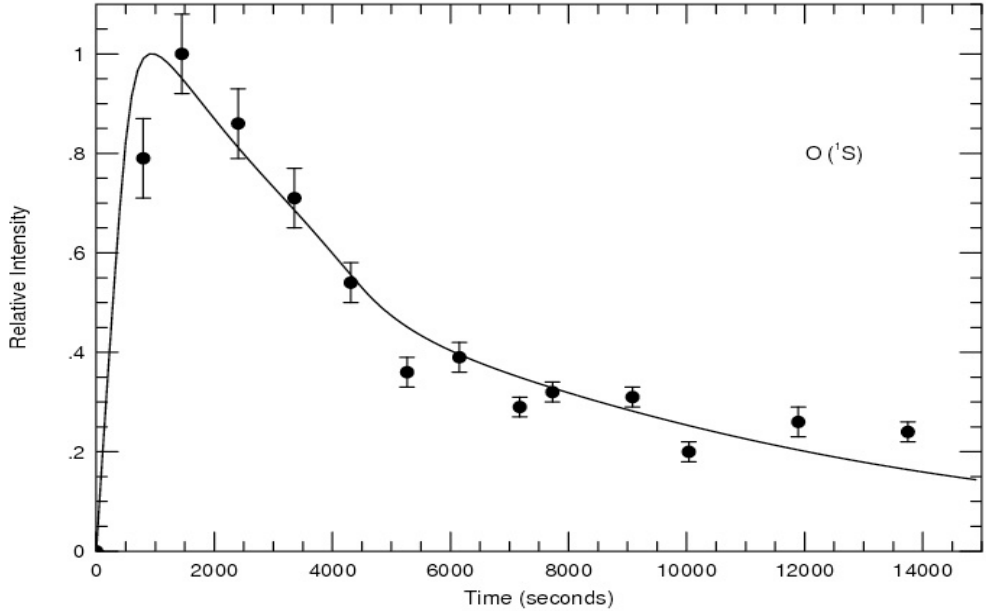

Figure 2. Least squares fit to the O($^1$S) data using a prompt and delayed three-step model. The O($^1$S) cometary emission was deconvoluted from the O($^1$S) of the upper atmosphere. The error bars are 3 σ and they were computed from the statistical error associated with total number of photons measured at each time. The relative weight of the prompt and delayed model was 0.375/0.635, respectively.

    The least squares fit to the data is shown in Fig. 2. Prompt and delayed three-step models with the same rate constants were used for the least squares fit. The rate constant for the decay of the O($^1$S) was fixed at 1.26 s$^{-1}$ and the rate constant for the decay of H$_2$O was fixed at 2.0 x 10$^{-5}$ s$^{-1}$. That is the value determined from the fit to the OH emission. It should be emphasized that since we are modeling the relative time response we do not require the branching ratio into a particular channel; we only require the overall time response for that parent. With these constraints the least squares fit yields $k_i$ = 2.2 x 10$^{-4}$ s$^{-1}$, $k_P$ = 3.3 x 10$^{-4}$ s$^{-1}$, [n$^0_p$]/[γ]=0.37/0.63 and a $\Delta^2$ =0.07. The $k_i$ indicates that the characteristic time for sublimation of water from the fresh ice surface is 4500 s. As previously mentioned the rate constant for the decay of the daughter, that is the emission of the green line in reaction 18, is very fast and it tends to dominate the mechanism (Ralchenko et al., 2008).

    O($^1$S)         →   O($^1$D)     +     hν$_{557.73}$                            (18)

The large value of $k_P$ derived from the least squares fit to the $O(^1S)$ emission is too big to be associated with the photodissociation of OH or any other oxygen-containing molecule. The only oxygen-containing molecule likely to have an appreciable abundance in comets that is known to produce $O(^1S)$ is $CO_2$ and its photodissociation rate constant is too slow.

This suggests that another mechanism is needed for the production of $O(^1S)$. Consider the following reaction,

$$H_2O + h\nu_{VUV} \rightarrow O(^3P) + 2H(^2S) \qquad (19)$$
$$H_2O + h\nu_{VUV} \rightarrow O(^1D) \text{ and/or } O(^1S) + H_2 \qquad (20)$$

Both of these reactions are known to occur when water is dissociated at Lyman alpha (Harich, et al. 2000) except that no laboratory experiments have yet shown that $O(^1S)$ is produced in reaction 20. The yields are small but this only affects the branching ratio and not the $k_G$. Some method other than direct photoexcitation is needed to produce the $O(^1S)$ from the $^3P$ and $^1D$ formed in reactions 19 and 20 because such a reaction would be too slow since both of them are optically forbidden. Electrons could be used to excite these transitions via the following reaction,

$$O(^3P) \text{ and / or } O(^1D) + e^- \rightarrow O(^1S) \qquad (21)$$

The rate constant determined from the modeling can be used to probe if reasonable electron excitation cross-sections and electron densities are consistent with it. Electrons with a minimum energy of 2.2 to 4.4 eV are needed to excite $O(^1D)$ and $O(^3P)$, respectively, to the $^1S$ state. The corresponding electron velocities are $8.7 \times 10^7$ to $1.2 \times 10^8$ cm sec$^{-1}$, respectively. The rate constant that is derived in the least squares fit is a pseudo first order rate constant i.e. $k_P = \sigma v \rho_e = 3.3 \times 10^{-4}$ s$^{-1}$. Using an electron density of $2 \times 10^4$ cm$^{-3}$ and the velocities associated with the minimum energy we can calculate the cross sections required for reaction 21 to fit the derived rate constant. The cross section for reaction 21 can be estimated to be $\sim 1.4$ to $1.9 \times 10^{-16}$ cm$^2$, which is certainly reasonable.

The delayed model is consistent with our knowledge that the green emission is observed in other comets and it supports the idea that it is formed from the photodissociation of water.

**CN**

The time dependences of the CN radical in Fig. 3 clearly show that there is a step in the curve that simply can not be fitted with a model that has a series of first order reactions without introducing a delay with a driving function in the mechanism as was done for $O(^1S)$. A prompt and delayed two-step model was used with the least squares program to fit the experimental data points that are shown with 5 σ error bars. The rate constants obtained from a least squares fit to the data used for the reactions forming the parent and the daughter in the prompt and delayed emissions are the same, namely $k_P = 0.14 \times 10^{-5}$ s$^{-1}$, $k_d = 3 \times 10^{-6}$ s$^{-1}$, and a $k_i = 8.3 \times 10^{-4}$ s$^{-1}$. The fit also yields a $[n^0_p]/[\gamma]$ equal to 0.25/0.75 and a $\Delta^2$ of 0.05, which is very good but not as good as the 5σ error bars.

To illustrate how the rate constants and the other values change if we accept larger values of $\Delta^2$, we allowed it to rise by a factor of 3 to 0.15. This drops the $k_D$ by a factor 3, raises the $k_P$ by a factor of 3, changes the $k_i$ to $1 \times 10^{-3}$ s$^{-1}$ and the ratio of $[n^0_p]/[\gamma]$ to 0.099/0.801.

The solar photodestruction rates at 1.51 AU (Crovisier, 1994) of the HCN, $C_2N_2$, $CH_3CN$, $HC_2CN$ and $NCC_2CN$ are $0.48 \times 10^{-5}$ s$^{-1}$, $1.4 \times 10^{-5}$ s$^{-1}$, $2.9 \times 10^{-5}$ s$^{-1}$, $1.2 \times 10^{-5}$ sec$^{-1}$, and $2.2 \times 10^{-5}$ s$^{-1}$, respectively. HCN photodestruction follows equation 22.

$$HCN + h\nu \rightarrow CN + H \qquad (22)$$

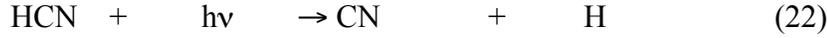

The photodissociation rate for HCN is closest to the $k_P$ obtained in the least squares fit but it is a factor of three lower. This lower value may arise from shielding of the sunlight by water and dust, which is not included in the model. The loss rate constant for CN according to Huebner et al.(1992) varies with solar activity between 1.3 and $3.1 \times 10^{-6}$ s$^{-1}$. This is in agreement with the rate from the fit of $3 \times 10^{-6}$ s$^{-1}$. The fact that in order to fit the observed response curve for the CN emission we have to invoke a prompt and a delayed process suggests that the impact released an amount of HCN and this was followed by a later release of HCN from the newly exposed surface. The rate of loss of HCN from the fresh ice is faster than it is for the parents of the other emissions as indicated by the larger value of the $k_i$. That may be a result of higher volatility of this molecule, which is consistent with a surface density, $\gamma$, larger than the bulk density, $n^0_p$, evacuated by the impact. Even though HCN appears to be the principal source of CN, it certainly does not preclude the presence of small amounts of other CN precursors.

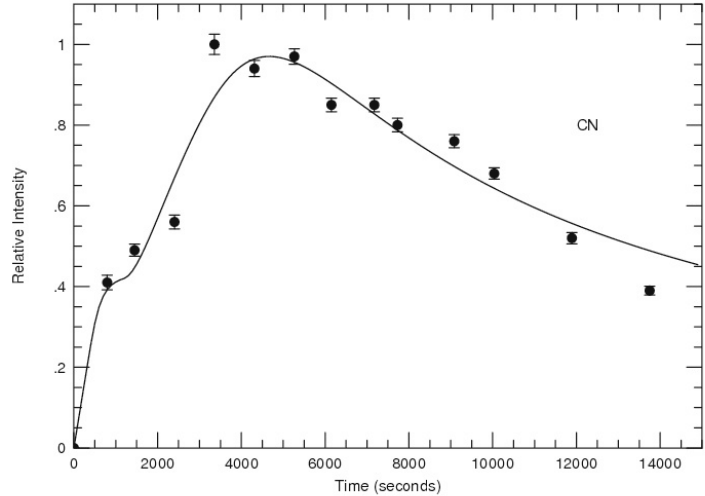

Figure 3. Least squares fit to the CN data using a prompt and delayed two-step model. The error bars are 5 $\sigma$ and they were computed from the statistical error associated with total number of photons measured at each time. The relative weight of the prompt and delayed model was 0.25/0.75, respectively.

## $C_2$

The time response of the $C_2$ Swan emission is shown in Fig. 4, where once again there is evidence for a prompt rise followed by a fall and then after a delay another rise. A variety of models were tried for fitting the time response subject to the criteria that any of the fits have to use the same rate constants for the production and loss of $C_2$ radicals in the prompt and delayed models. The fit shown is one that consists of a prompt two-step model followed by a delayed three step model using these two constraints. These constraints are equivalent to assuming that the parent and the loss mechanisms for $C_2$ radicals are the same in the prompt and the delayed model. The later constraint almost has to be true since the $C_2$ radical is involved in both cases. The most likely parent for $C_2$ is $C_2H$, which in the three step model is produced by photodissociation of $C_2H_2$ (Jackson 1976). The reactions summarize the two step mechanism for the prompt dissociation:

$$C_2H + h\nu \rightarrow C_2 + H \quad (23)$$
$$C_2 + h\nu \rightarrow C_2^+ + e^- \text{ or } 2C \quad (24)$$

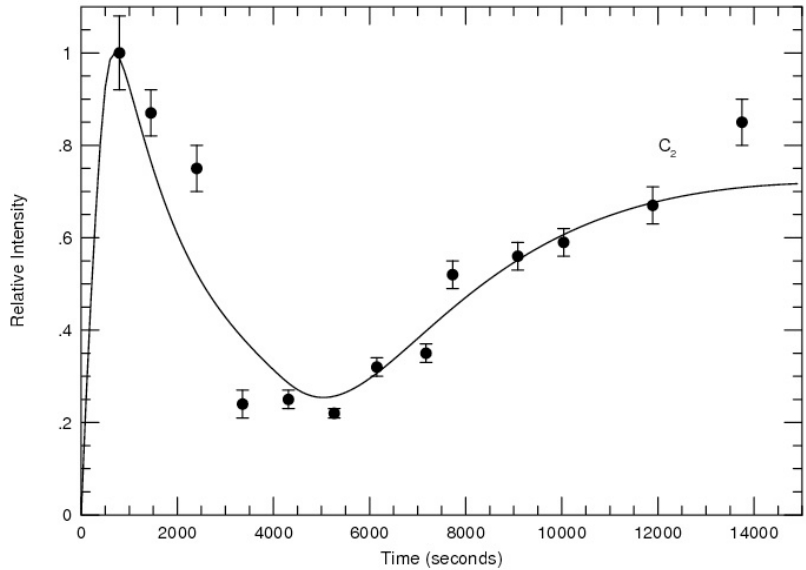

Figure 4. Least squares fit to the $C_2$ data using a prompt two-step model and a delayed three-step model. The error bars are 3 σ and they were computed from the statistical error associated with total number of photons measured at each time. The fit assumes that the branching ratio for the grandparent to parent in the three step model is one. See the discussion. The relative weight of the prompt and delayed model was 0.001/0.999, respectively.

In the delayed three-step model these two reactions are preceded by the following reaction for the production of $C_2H$:

$$C_2H_2 + h\nu \rightarrow C_2H + H \qquad (25)$$

The three-step mechanism for the delayed $C_2$ emission agrees with the previously proposed mechanism used to explain $C_2$ in comets (Jackson, 1976). This mechanism can and does produce $C_2$ radicals in a variety of electronic states as both laboratory (Jackson et al., 1978; McDonald et al., 1978; Urdahl et al., 1988; Urdahl et al., 1989; Bao et al., 1991), observational (Sorkhabi et al., 1997), and theoretical (Mebel et al., 2001; Apaydin et al., 2004) studies have shown.

The best least squares fit to the observational data with the constraints that parent and daughter rate constants are the same yielded values for the rate constants of $k_g = 5.0 \times 10^{-6}$ s$^{-1}$, $k_p = 1.4 \times 10^{-3}$ s$^{-1}$, $k_d = 1.5 \times 10^{-5}$ s$^{-1}$, and $k_i = 2.5 \times 10^{-4}$ s$^{-1}$ and a value for $[n^0_p]/[\gamma]$ of 0.01/0.99. The curve corresponding to this fit to the data is shown in Fig. 4. The error bars in this figure are $3\sigma$ and the $\Delta^2$ for the fit is 0.11. The solar photodissociation rate constants for acetylene at 1 AU are reported to be between 2.0 and $20 \times 10^{-5}$ s$^{-1}$, which at 1.51 AU will be 0.88 to $8.8 \times 10^{-5}$ s$^{-1}$ (Crovisier 1994). The rate constant for the grandparent used in the delayed three-step model is slightly smaller than the smallest of these rate constants, suggesting again that the opacity of the coma needs to be taken into account in the model. Lower and higher $k_g$ lead to poorer fits to the data points. The rate constant for the depletion of the parent in the ice is in the range of most of the others seen in this study. The use of a two step model followed by a three step model does introduce the branching ratio for the grandparent into the model since this will not be divided out when the relative time dependence is derived. In the present case this is not such a problem because all of the laboratory evidence suggests this is very close to one (Jackson, 1974).

The rate constant for the photodissociation of $C_2H$ in reaction 23 has not been measured with the accuracy of those for acetylene. It should be at least as large as the rate constant for the photodissociation of acetylene because removal of the H atom should shift the absorption to longer wavelengths, as it does for most free radicals. The present results support this point of view because the rate of dissociation of the "parent", i.e., $C_2H$, is larger than the rate for photodissociation of $C_2H_2$. Jackson et al.(1996) and Heubner et al. (1992) both tried to estimate this rate constant from the theoretical calculations and their values of $0.7 \times 10^{-5}$ and $0.01 \times 10^{-5}$ s$^{-1}$, respectively, are much lower than the value derived from the fit to the observations. A larger value of $6 \times 10^{-5}$ s$^{-1}$ can be calculated from the approximate absorption cross section for $C_2H$ measured by Fahr (2003) from 235-261 nm. Within this limited wavelength range it is likely that the absorption cross section, and hence this rate constant, is a lower limit of the true value for the rate constant for the photodissociation cross section for $C_2H$. Even so it is likely that other types of reactions need to be considered, like the interaction of electrons, as a means of increasing the loss rate of $C_2H$.

## $C_3$

The time response of the $C_3$ emission derived from the Keck 1 HIRES observations of the Deep Impact encounter is shown as points in Fig. 5 with the $3\sigma$ error bars. The curve in Fig. 5 is a least square fit to the data using a three-step model with the same rate constants for the prompt and delayed emission. From the fit one derives rate constants for the grandparent, parent, and daughter of $8.0 \times 10^{-5}$ s$^{-1}$, $4.0 \times 10^{-3}$ s$^{-1}$ and $7.5 \times 10^{-5}$ s$^{-1}$, respectively. The other parameters determined by the fit are $2.9 \times 10^{-4}$ s$^{-1}$ for the $k_i$, 0.07 for the $\Delta^2$, and 0.07/0.93 for the $[n^0_p]/[\gamma]$. The depletion rate constant is similar to the value of the others that have been determined in this study indicating similar thermal properties. Again the surface density is considerably higher than the bulk density from the evacuation because of the impact.

Stief (1972) first suggested that this radical was produced by the vacuum ultraviolet photodissociation of propyne, $CH_3C_2H$, in which the radical was produced by the sequential loss of two $H_2$ molecules.

$$CH_3C_2H + h\nu \rightarrow HCC_2H + H_2 \quad (26)$$
$$HCC_2H + h\nu \rightarrow C_3 + H_2 \quad (27)$$

This would lead to a two-step model that is not in agreement with the present data or other studies. As Jackson (1976) suggested, it is more likely that $C_3$ is produced by a three-step reaction involving propyne, $CH_3C_2H$. Later work suggested that the isomer of propyne, i.e., allene ($H_2C_3H_2$), could also produce the intermediate $C_3H_2$ via reaction 28 and by absorbing a second photon produce $C_3$ via reaction.

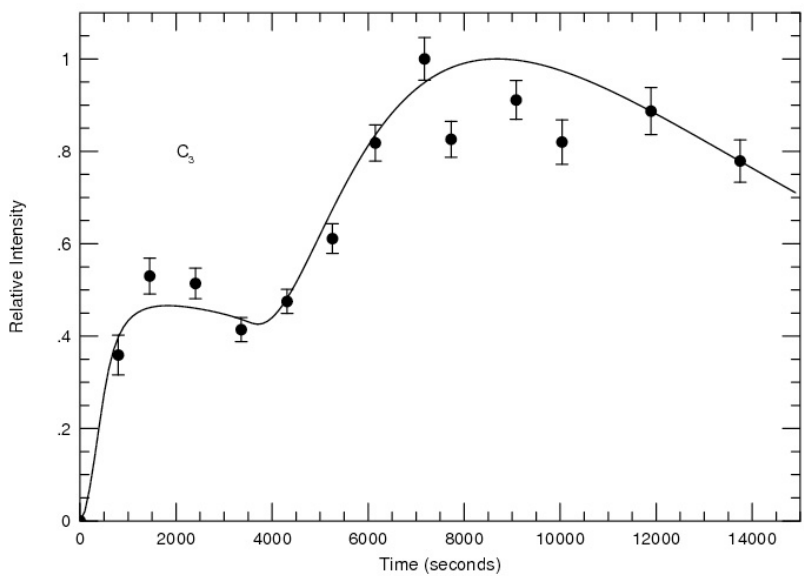

Figure 5. Least squares fit to the $C_3$ data using a prompt and delayed three-step model. The error bars are 3 $\sigma$ and they were computed from the statistical error associated with total number of photons measured at each time. The relative weight of the prompt and delayed model was 0.074/0.926, respectively.

$$H_2C_3H_2 + h\nu \rightarrow C_3H_2 + H_2 \quad (28)$$
$$C_3H_2 + h\nu \rightarrow C_3 + H_2 \quad (29)$$

In this mechanism, allene or propyne are the grandparents and the intermediate, $C_3H_2$ radical is the parent that produces a $C_3$. Indeed laser induced fluorescence (LIF) and photofragment spectroscopy studies showed that this radical could be formed in this manner when an ArF laser is used as the photolysis source (Jackson et al.1991; Song et al. 1994). These studies also showed that allene, $H_2C=C=CH_2$, was 8 times more efficient in producing this reaction at this wavelength, and, when this is coupled with the fact that the photochemical lifetime is shorter because the solar absorption is shifted to longer wavelengths, it suggests that this may be the mechanism of choice. $C_3$ is destroyed via equation 30:

$$C_3 + h\nu \rightarrow C_2 + C \quad (30)$$

High quality theoretical studies have been done to explain this mechanism and show that the dissociation occurs on the ground state surface after a rapid internal conversion process. It is easier for allene rather than propyne to do this (Jackson et al.1998). While this works at 193.3 nm it does not tell us what happens to these molecules when exposed to the Sun. The photodestruction rate constants for allene and propyne have been reported to be $\sim 6 \times 10^{-5}$ s$^{-1}$ at 1.51 au, respectively (Helbert et al.2005). This is slightly slower than the rate constants determined from the fit and suggests that the model may need an additional source of the parent intermediate such as electron collisions with a $C_3H_4$ grand parent. The destruction of the $C_3$ via reaction 30 has been reported to have a rate constant that varies from $0.9 \times 10^{-5}$ by (Helbert et al.2005) to $4 \times 10^{-5}$ by (Heubner 1992) at 1.51 au. This is certainly in the range of the rate constant derived for this reaction in the present study.

**CH**

The $CH (A^2\Delta) \rightarrow CH (X^2\Pi)$ emission is observed in many comets, and, extracting the time dependence from the Keck emission spectrum using the procedure that has been described yields the points shown in Fig. 6 with $3\sigma$ error bars. The least squares fit to the points was accomplished using a prompt and delayed three-step model employing the same rate constants for both models. The least square fit has a $\Delta^2$ of 0.06 and 0.16/0.84 for $[n^0_p]/[\gamma]$. The value of $\Delta^2$ is consistent with the error and the surface density is much greater than the bulk density of the evacuated material. The fit is constrained by fixing the rate constant for the loss of CH. This rate constant is well known and it is very fast at $1.3 \times 10^{-2}$ s$^{-1}$ at 1 au (Huebner et al. 1992). At the heliocentric distance of the Deep Impact encounter of 1.51 au this rate constant becomes $6 \times 10^{-3}$ s$^{-1}$. The least squares fit to the data results in values for $k_G$, $k_P$, and $k_i$ of $0.1 \times 10^{-6}$ s$^{-1}$, $2.0 \times 10^{-4}$ s$^{-1}$, and $2.5 \times 10^{-4}$ s$^{-1}$, respectively. The $k_i$ is similar to the others that have been derived from the least squares fit. The rate constant for the parent is fast and suggests that the parent might be a free radical because their absorption spectra are shifted to the red. This will increase the photodestruction rate constants because the solar flux is higher than it is at shorter wavelengths. Further, the bond energies of free radicals are generally weaker, so they will have a threshold for dissociation at longer wavelengths.

A likely free radical that could be a parent for CH is $CH_2$. There are two molecules, namely, methane and ethane, that laboratory evidence suggests could be the grandparent to produce this parent. Both methane and ethane molecules have been observed in comets using the Keck II- NIRSPEC instrument by Mumma et al. (2005). The rate constant for the grandparent is smaller than the rate constants of $4 \times 10^{-6}$ s$^{-1}$ for methane at 1.51 au. This can be explained by the opacity of the cloud at Lyman $\alpha$, the wavelength for the photolysis of methane and ethane. Mordant et al.(1993) have studied the photodissociation of methane and suggested that one of the primary processes produces CH(X $^2\Pi$) directly via reaction 31. Subsequent work by several authors has shown that the principal reactions produce $CH_2(a^1A_1)$ in reaction 32 (Cook et al.2001; Wang et al.2000; Heck et al.1996).

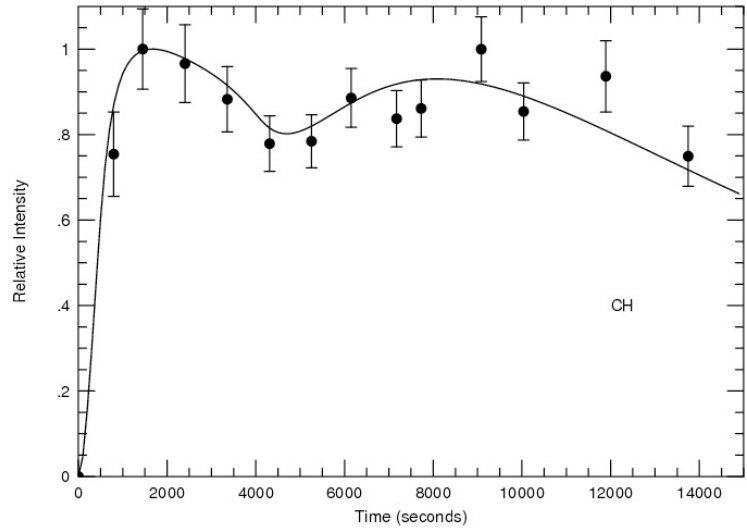

Figure 6. Least squares fit to the CH data using a prompt and delayed three-step model. The error bars are 3 $\sigma$ and they were computed from the statistical error associated with total number of photons measured at each time. The relative weight of the prompt and delayed model was 0.16/0.84, respectively.

$$CH_4 \;+\; h\nu \;\rightarrow\; CH(X^2\Pi) \;+\; H \;+\; H_2 \quad (31)$$
$$CH_4 \;+\; h\nu \;\rightarrow\; CH_2(a^1A_1) \;+\; H \;+\; H_2 \quad (32)$$

Ethane is also a candidate for producing $CH_2$ via reaction 33, as suggested in earlier work on the VUV photolysis of $C_2H_6$ (Hampson and McNesby, 1965).

$$C_2H_6 \;+\; h\nu \;\rightarrow\; CH_2 \;+\; CH_4 \quad (33)$$

It may be that both methane and ethane are contributing to the observed CH emission during Deep Impact. The fact that the two three step models used to fit the observed data employ the same rate constants implies that the same molecules are involved in the direct and delayed emission.

**NH and NH$_2$**

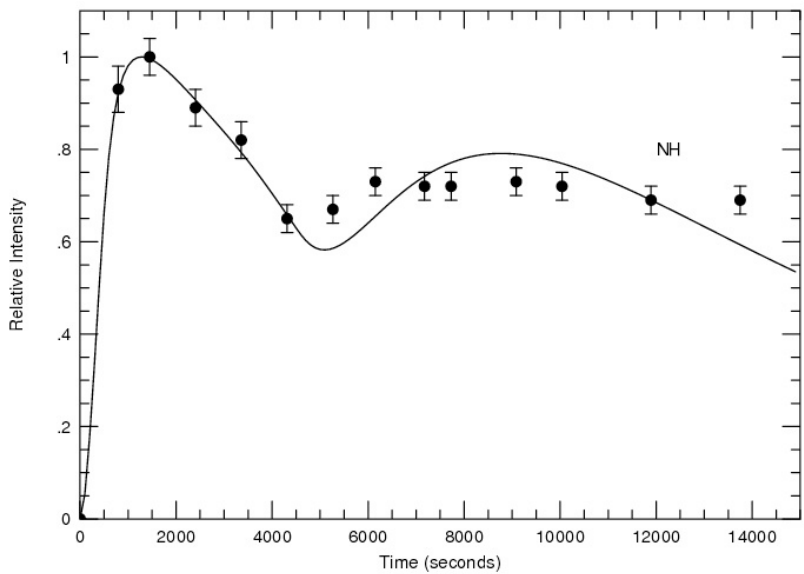

Figure 7. Least squares fit to the NH data using a prompt and delayed three-step model. The error bars are 3 σ and they were computed from the statistical error associated with total number of photons measured at each time. The relative weight of the prompt and delayed model was 0.12/0.88, respectively.

Ammonia is thought to form NH via a three step mechanism involving first the photodissociation via reaction 34 to NH$_2$ + H. This is then followed by the photodissociation of NH$_2$ via reaction 35 to form NH, which then undergoes photodissociation via reaction 36.

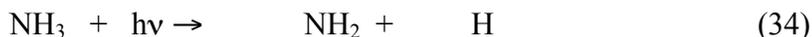
$$NH_3 + h\nu \rightarrow NH_2 + H \qquad (34)$$

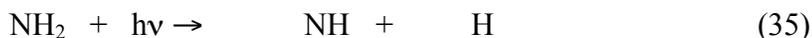
$$NH_2 + h\nu \rightarrow NH + H \qquad (35)$$

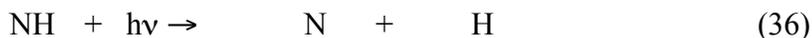
$$NH + h\nu \rightarrow N + H \qquad (36)$$

This puts severe constraints on the models that are used to fit the NH and NH$_2$ data in Figs. 7 and 8. It requires that if a three step model is used to fit the NH data then a two-step model must be used to fit the NH$_2$ data and that the rate constant used for the parent in the three step model for NH is equal to the one used for the two step model for NH$_2$ since they are the same molecule. The rate constant for reaction 36 is determined by the solar photodissociation of the NH free radical. The rate constant for the photodissociation of the NH is well known since it is based upon laboratory measurements of the radiative lifetime of the NH radical. Singh and Gruenwald (1987) have calculated a value for the photodissociation rate constant at 1 au of 5 x 10$^{-5}$ s$^{-1}$ from the laboratory work, which at 1.51 au is 2.0 x 10$^{-5}$ s$^{-1}$. The least squares fit to the NH data in Fig. 7 then gives the rate constants $k_G$ = 6.5 x 10$^{-3}$, $k_P$ = 3.5 x 10$^{-4}$ s$^{-1}$, $k_d$ = 2.0 x 10$^{-5}$ s$^{-1}$ and $k_i$ = 2.2 x 10$^{-4}$ s$^{-1}$ with a $\Delta^2 = 0.03$ and $0.12/0.88$ for the $[n^0_p]/[\gamma]$. Using the daughter rate constants for NH, like OH, as well as the additional constraint for $k_P$ should provide excellent constraints for the rest of the model. The value is of the same order of magnitude as the OH radical, as it should be since the radiative lifetimes and absorption regions where the predissociation occurs are similar. The rate constant for dissociation of the grandparent via reaction 34 obtained in the modeling is $6.5 \times 10^{-3}$ s$^{-1}$. The literature values for the photodissociation of NH$_3$ at 1.51 au range from $6.6 \times 10^{-5}$ s$^{-1}$ to $2.1 \times 10^{-4}$ s$^{-1}$, which is much slower than the value derived in the modeling. This may be a reflection that the present model does not contain electron molecule interactions or additional parents such as hydrazine, the propellant for the rocket motors.

Figure 8 shows the least squares fit to the NH$_2$ data using the constraints discussed above. The rate constant derived from the fit to the data in Figs 7 and 8 for the photodissociation of NH$_2$ are several orders of magnitude larger than the rate constants reported by Heubner et al. (1992). Their rate constants vary from $2.2 \times 10^{-6}$ s$^{-1}$ to $3.4 \times 10^{-6}$ s$^{-1}$ for the quiet and active suns, respectively. They are based upon the theoretical cross sections calculated by Saxon (1983). These neglect predissociation and as a result under-estimate the true rate constant. The methods used in the earlier rate calculations are more approximate than the ones that are available today and newer calculations are needed. Recent calculations of the transition probabilities of NH$_2$ by Vetter et al. (1996) can be used to reevaluate the photodissociation rate constant. Their work suggests that the photodissociation occurs in the 163 nm region where the solar flux has decreased. This implies that the rate constant will be relatively small. To obtain a larger rate constant for the loss of $NH_2$ will probably require the inclusion of collisions with electrons, similar to what was required for $O(^1S)$.

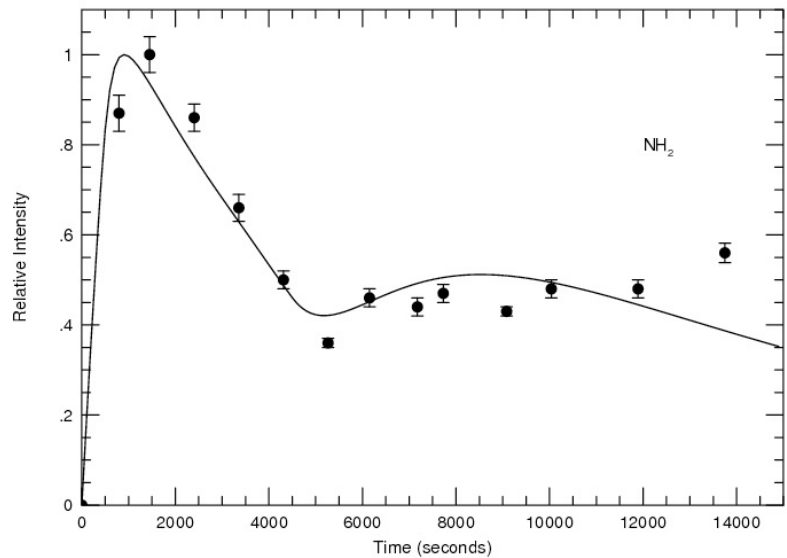

Figure 8. Least squares fit to the NH$_2$ data using a prompt and delayed three-step model. The error bars are 3 σ and they were computed from the statistical error associated with total number of photons measured at each time. The relative weight of the prompt and delayed model was 0.0002/0.9998, respectively.

The two step models used in the least squares fit to the $NH_2$ data in Fig. 8 yields different $k_P$'s of $1.1 \times 10^{-4}$ s$^{-1}$ and $1.0 \times 10^{-7}$ s$^{-1}$ for the prompt and delayed emissions, respectively. The fit also yields a $\Delta^2 = 0.07$ and a 0.0002/0.9998 for the $[n^0_p]/[\gamma]$. This implies that there are either two different parents for $NH_2$ or, if the parent is the same, there must be two different kinds of reactions involved in the prompt and delayed emission. The least squares fit also yields a rate constant for the 1/e value of the emission from the fresh ice, $k_i = 2.2 \times 10^{-4}$ s$^{-1}$, that is the same as the value obtained for NH. The $k_P$ for the prompt emission is nearer to the lower end of the literature values for the photodissociation of $NH_3$ at 1.51 au but the delayed rate constant is several orders of magnitude slower. If $NH_3$ is the parent for both the prompt and delayed emissions then the model has to be modified in a way that effectively lowers the first order decay constant for the parent at longer times. If the first order decay constant is due to photodissociation it implies that less light is reaching the ammonia, the absorption coefficient decreases or the absorption wavelengths shift to the blue. Shifting the absorption wavelength to the blue is consistent with the parent for the delayed emission being the $NH_3$-$H_2O$ complex instead of $NH_3$. In this case the delayed emission would involve the following reaction,

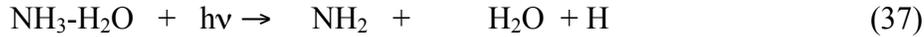
$$NH_3\text{-}H_2O \ + \ h\nu \rightarrow \ NH_2 \ + \ H_2O + H \tag{37}$$

Recent high quality theoretical calculations of this water-ammonia complex show that the absorption is shifted 0.5 eV to the blue and the absorption coefficient increases by 20% (Lane et al.2008). This increase in the absorption is probably not enough to compensate for the decrease in the solar flux at shorter wavelengths. A more detailed model is required to evaluate this possibility.

## 5. VARIATION OF THE NUCLEAR COMPOSITION

A summary of the data derived from modeling the changes in the temporal responses in the emissions following impact of the projectile during the Deep Impact mission is given in Table 1. The results for the $[n^0_p]/[\gamma]$ in the table answers one of the principal goals of the Deep Impact mission: to determine whether the chemical composition of the surface was identical to the composition of the interior of the comet. Our modeling of the time response of the radical emissions shows that there are distinct changes in the ratio of the concentration of the composition of the gas released in the collision, $[n^0_p]$, and the density of the fresh surface ice, $[\gamma]$. The density of the fresh surface ice is higher for all of the emissions except $O(^1S)$. In the model for this emission the parent is $H_2O$. It should not change since it is the glue that is holding the dust and other gases together. Water is unique with a very high latent heat of vaporization that controls the sublimation of all of the gases and dust in comets. The fact that there is a large change in the $[n^0_p]/[\gamma]$ for the other emissions implies that in the comet nuclei there is a gradient of the parent gases as one goes from the surface of the nucleus into the interior. These are not minor changes since they vary from a factor of three for the parent of HCN to a factor of 5000 for the parent of $NH_2$. It clearly indicates that measuring the composition of the gases emitted from the surface of the ice does not provide one with a complete picture of the chemical composition of the comet. Thus, to connect the composition of the comet to the chemical composition of the early solar system requires experiments that can determine sub-surface chemical composition. Without this capability, any such measurements will only be measuring compositions that have been modified by repeated passages around the sun.

The question of the variation of a comet's composition with depth has been approached in the past with studies of split comets. The split in 1995 and the further split in 2006 of Comet 73P/Schwassmann-Wachmann 3 (SW3) is an example of this. In addition, in 2006, SW3 approached close to the Earth, making it suitable to compare the composition of different fragments. This, then, was a natural experiment to view the inside of a comet. The fragments all showed the same composition (e.g. Dello Russo et al. 2007). This would imply that the comet had a uniform composition, in contrast with what we observed from our Deep Impact HIRES data. However, these observations may not be in disagreement because of the nature of the two events. With Deep Impact, we were watching the spectrum evolve in the first few hours after an impulse, with no time for the coma to equilibrate. In addition, we were able to accurately remove the underlying ambient spectrum and just determine the delta composition. In addition, we had extremely good spatial resolution so we could follow the progress of the gas. Thus, the Deep Impact experiment coupled with the Keck HIRES observations allowed for a detailed look that is not possible in any other way.

## 6. CONCLUSIONS

The Keck-HIRES measurements of the Deep Impact encounter uncovered new information about the cometary nucleus and the chemical processes occurring in comets. Modeling the data has shown that there is a concentration gradient in the cometary nucleus. This confirms that the upper layers of the comet are depleted in the parents and grandparents that form the emissions that are observed in comets. This has been suspected for some time and is included in some of the detailed models of the nucleus.

The rate constants derived from the least squares fits to the emissions of OH, CN, $C_3$ and $C_2$ are consistent with the solar photodissociation rate constants for $H_2O$, HCN, $HC_2H$ and $H_2CCH_2$ or $CH_3C_2H$. The first two emissions are fit with a two-step model that confirms the rate constants for solar photodissociation. HOH and HCN produces OH and CN even though the quality of the least squares fit is not as good as the 5 σ error bars of the data. This suggests that the Haser model used in the least squares fit is not detailed enough to completely fit the data. A three-step model with $H_2CCH_2$ or $CH_3C_2H$ was needed to fit the $C_3$ emission as had been suggested earlier by Stief (1972) and Jackson (1976) and the rate constant derived for the $k_G$ is consistent with the rate constants expected for solar photodissociation. Similarly, a three-step model is needed to explain the delayed emission for $C_2$ but only a two-step model could be used for prompt emission. This has led us to suggest that this is due to direct production of the parent of $C_2$, i.e., $C_2H$, during the impact.

Some of the rate constants derived for the $k_P$ for $O(^1S)$, $C_2$, $C_3$, CH, NH and $NH_2$ are much larger than one can explain by simple photodissociation. This suggests that there is an additional reaction that has to be invoked to explain these reactions. Modeling the results for the emission of the O($^1$S) provides a clue to this reaction. Collisions of low energy electrons with O($^3$P) appear to be fast enough to explain these results. Similar collisions can be invoked to excite molecules and free radicals to states that dissociate to the observed products.

The method that we used is more accurate than the light curve method for studying the impulsive activity after the impact because we have removed the ambient background. A least squares fit to the data with the constraints determined by laboratory measurements as well as the consistency in the model between the observed species is used to fit the data points for time dependence of the radical emissions after Deep Impact. By employing constraints and consistency, we limit the parameter space available for the least squares fit and simultaneously insure that the emissions from related species such as NH and NH$_2$ are consistent. The data that we have extracted from the Deep Impact observations have a very well defined time scale. Our analysis does, however, require knowledge of the flow rate of the gas. The error bars in the observations are random errors derived from the number of observed photon counts and do not include systematic errors which may be present. Thus, they are certainly an underestimate of the true errors. The square of the residuals between the data and the model has been used to provide an objective criterion for determining the quality of the fit. Changing the rate constants by a factor of two tends to change the value of the $\Delta^2$ by a similar amount. Within the constraints described in the modeling the rate constants derived in this paper are probably accurate within a factor of two.

The great light gathering power of the Keck telescope and its excellent spatial resolution along with the impulsive nature of the impact and our precise knowledge of the time of the impulse has allowed us to extract unique information from the Deep Impact encounter. This in turn has provided us new insights into the structure of the cometary nucleus. Still there are questions that have been raised that can only be solved by employing more elaborate models such as ComChem (Helbert et al.2005) to fit the data and with new data on the collisions of electrons with some of the parents of the observed radicals. We plan to employ this model in our future work using the data set that has been extracted from the observations.

WMJ, XLY and XS were supported by NSF grant 0503765. ALC was supported by a NASA Grant NNG04G162G. The authors wish to recognize and acknowledge the very significant cultural role and reverence that the summit of Mauna Kea has always had within the indigenous Hawaiian community. We are most fortunate to have the opportunity to conduct observations from this mountain.

# Table 1
Rate constants derived from the least squares fits to the temporal response of the radical emissions observed from collision of Deep Impact with Comet 9P/ Tempel. The rate constants refer to the comets heliocentric distance of 1.51 au at the time of the collision.

| Radical | Model | $[n^0_p]/[\gamma]$ | $\Delta^2$ | $k_i$ (s$^{-1}$) | $k_G$ (s$^{-1}$) | $k_P$ (s$^{-1}$) | $k_D$ (s$^{-1}$) |
|---|---|---|---|---|---|---|---|
| **OH** | 2 step | na | 0.13 | na | na | $2.0 \times 10^{-5}$ | $4.0 \times 10^{-6}$ |
| **O($^1$S)** | 3 step | 0.37/0.63 | 0.07 | na | $2.0 \times 10^{-5}$ | $4.0 \times 10^{-4}$ | 1.26 |
| | 3 step | 0.37/0.63 | 0.07 | $2.2 \times 10^{-4}$ | $2.0 \times 10^{-5}$ | $4.0 \times 10^{-4}$ | 1.26 |
| **CN** | 2 step | 0.25/0.75 | 0.05 | na | na | $1.4 \times 10^{-6}$ | $3.0 \times 10^{-6}$ |
| | 2 step | 0.25/0.75 | 0.05 | $8.3 \times 10^{-4}$ | na | $1.4 \times 10^{-6}$ | $3.0 \times 10^{-6}$ |
| **C$_2$** | 2 step | 0.001/0.999 | 0.11 | na | na | $1.4 \times 10^{-3}$ | $1.5 \times 10^{-5}$ |
| | 3 step | 0.001/0.999 | 0.11 | $2.5 \times 10^{-4}$ | $5.0 \times 10^{-6}$ | $1.4 \times 10^{-3}$ | $1.5 \times 10^{-5}$ |
| **C$_3$** | 3 step | 0.074/0.926 | 0.073 | na | $8.0 \times 10^{-5}$ | $4.0 \times 10^{-3}$ | $7.5 \times 10^{-5}$ |
| | 3 step | 0.074/0.926 | 0.073 | $2.9 \times 10^{-4}$ | $8.0 \times 10^{-5}$ | $4.0 \times 10^{-3}$ | $7.5 \times 10^{-5}$ |
| **CH** | 3 step | 0.16/0.84 | 0.055 | na | $1.0 \times 10^{-7}$ | $2.0 \times 10^{-4}$ | $6.0 \times 10^{-3}$ |
| | 3 step | 0.16/0.84 | 0.055 | $2.5 \times 10^{-4}$ | $1.0 \times 10^{-7}$ | $2.0 \times 10^{-4}$ | $6.0 \times 10^{-3}$ |
| **NH** | 3 step | 0.12/0.88 | 0.03 | na | $6.5 \times 10^{-3}$ | $3.5 \times 10^{-4}$ | $2.0 \times 10^{-5}$ |
| | 3 step | 0.12/0.88 | 0.03 | $2.2 \times 10^{-4}$ | $6.5 \times 10^{-3}$ | $3.5 \times 10^{-4}$ | $2.0 \times 10^{-5}$ |
| **NH$_2$** | 2 step | 0.0002/0.9998 | 0.07 | na | na | $1.1 \times 10^{-4}$ | $3.5 \times 10^{-4}$ |
| | 2 step | 0.0002/0.9998 | 0.07 | $2.2 \times 10^{-4}$ | na | $1.0 \times 10^{-7}$ | $3.5 \times 10^{-4}$ |